\documentstyle[12pt,psfig,aps]{revtex}

\begin{document}
\pagenumbering{arabic}
\title{\bf  Vacuum discharge as a possible source of gamma-ray bursts}
\author{G.~Mao, S.~Chiba}  
\address{Japan Atomic Energy Research Institute \\
Tokai, Naka, Ibaraki 319-1195, Japan}                              
\author{W.~Greiner}
\address{Institut f\"{u}r Theoretische Physik der 
J. W. Goethe-Universit\"{a}t\\
 Postfach 11 19 32, D-60054 Frankfurt am Main, Germany}
\author{K.~Oyamatsu}
\address{Faculty of Studies on Contemporary Society \\
Aichi Shukutoku University \\
9 Katahira, Nagakute, Aichi 480-1197, Japan}
\date{\today}
\maketitle
\begin{abstract}
\begin{sloppypar}
We propose that spontaneous particle--anti-particle pair creations from the
discharged vacuum caused by the strong interactions in dense matter are  
major sources of $\gamma$-ray bursts. Two neutron star collisions or 
black hole-neutron star mergers at cosmological distance could produce
a compact object with its density exceeding the critical density for
pair creations. The emitted anti-particles  annihilate with corresponding
particles at the ambient medium. This releases a large amount of energy.
We discuss the spontaneous $p\bar{p}$ pair creations within two neutron star
collision and estimate the exploded energy from $p\bar{p}$ annihilation 
processes. The total energy could be around $10^{51} - 10^{53}$ erg depending
on the impact parameter of colliding neutron stars. This value fits well into
the range of the initial energy of the most energetic $\gamma$-ray bursts.

\end{sloppypar}
\bigskip
\noindent {\bf PACS} number(s): 98.70.Rz; 52.80.Vp; 26.60.+c             
\end{abstract}
\newcounter{cms}
\setlength{\unitlength}{1mm}
\newpage
\begin{sloppypar}
Gamma-ray bursts (GRBs) were discovered accidentally in the late 1960s by
the Vela satellites. The discovery was announced in 1973 \cite{Kle73}. 
Since then,
they have been one of the greatest mysteries in high-energy astrophysics
for almost 30 years. The situation has improved dramatically in 1997, when
the BeppoSAX satellite discovered X-ray afterglow \cite{Cos97}, which
enabled accurate position determination and the discovery of optical 
\cite{Par97} and radio \cite{Fra97} afterglows and host galaxies. The 
distance scale to GRBs was finally unambiguously determined: their sources
are at cosmological distances \cite{Met97}. In spite of all these recent 
progress, we still do not know what produces GRBs! The nature of the 
underlying physical mechanism that powers these sources remains unclear.
The optical identification  and measurement of redshifts for GRBs allow us to
determine their distances and the amount of energy that would be radiated
in an isotropic explosion. In recent three observations 
(GRB971214 \cite{Kul98}, 980703 \cite{Djo98} and 990123 \cite{Kul99}),
the total isotropic energy radiated was estimated to be in excess of 
$10^{53}$ erg. For GRB990123, the inferred isotropic energy release is up 
to $3.4 \times 10^{54}$ erg, or $1.9$ $M_{\odot}$ (where $M_{\odot}$
is the solar mass), which is larger 
than the rest mass of  most neutron stars. 
It has been suggested that the explosion
of GRB990123 is not isotropic, which reduces the energy released in 
$\gamma$-rays alone  to be $6 \times 10^{52}$ erg \cite{Kul99} due to
finite beaming angle. However, if one adopts the picture of the 
fireball internal shock model \cite{Pir99} that random internal collisions
among shells produce the highly variable $\gamma$-ray burst emissions, the
required initial energy will be raised by a factor of about 100 since 
it is argued  that only
1\% of the energy of the initial explosion can be converted into the 
observed radiation \cite{Kum99}. Therefore, it appears that the total 
exploded energy for the most energetic bursts is close to or possibly greater
than $10^{54}$ erg.  It seems to be difficult to imagine a source that
could provide so much energy. The first and foremost open question 
concerning GRBs is what are the inner engines that power GRBs \cite{Pir99} ?

On the other hand, the GRB spectrum is nonthermal. In most cases there is
a strong power law high-energy tail extending to a few GeV. A particular
high-energy tail up to 18 GeV has been reported in GRB940217 \cite{Hur94}.
This nonthermal spectrum provides an important clue to
the nature of GRBs.

Various GRB models have been suggested in the literature, see e.g. Refs. 
\cite{Pir99,Eic89,Ruf97,Che96,Far99,Moh99,Woo99}. 
Among them, the neutron star merger
seems to be the most promising candidate. Three-dimensional hydrodynamical
simulations of the coalescence of binary neutron stars 
(NS-NS) \cite{Shi93,Ras94,Ruf97,Jan96},
direct collision of two neutron stars \cite{Ras92,Cen93,Ruf98} as well as 
black hole-neutron star (BH-NS) merger \cite{Jan99} 
have been performed by some authors.
The largest energy deposition of $ \sim 10^{51}$ erg by
$\nu\bar{\nu}$ annihilation was obtained in the black hole-neutron star
merger (for NS-NS collision, the total energy is around $10^{50}$ erg
 \cite{Ruf98}). 
This may account for certain low-energy GRBs on the one hand, but it
is, on the other hand, still far away from the energetic ones mentioned above.
However, it should be pointed out that in those macroscopic
simulations (and almost all GRB fireball models)
the effects of strong interactions, e.g., the modification of hadron properties in dense matter, many body effects, vacuum correlations et al.,  
have been largely neglected except that
a nuclear equation of state is applied.

In this Letter we propose an alternative scenario for the source of the
most energetic $\gamma$-ray bursts. It is well known that the density is 
fairly high at the center of neutron stars. The central density can be
several times nuclear saturation density \cite{Gle85}. 
Furthermore, superdense matter
could be formed at NS-NS/BH-NS mergers and direct NS-NS collisions. 
Three-dimensional hydrodynamical simulations showed that when two neutron
stars collide with a free-fall velocity, 
the maximum density of the compressed core can be 
1.4 (off-center collision, the impact parameter $b=R$, i.e., one neutron
star radius) to 1.9 (head-on collision) times the central density of a
single neutron star \cite{Ruf98}. At such high density,
not only the properties of baryons will be modified drastically 
according to the investigation of relativistic mean-field theory (RMF)
and relativistic Hartree approach (RHA) \cite{Ser86}, but also the
vacuum, i.e., the lower Dirac sea, might be distorted substantially 
\cite{Gre95} since the meson fields, which describe the strong interactions
between baryons, are very large. At certain densities, when the threshold
energy of the ``negative-energy sea''-nucleons (i.e., the nucleons
in the Dirac sea) 
is larger than the nucleon free mass,
the nucleon--anti-nucleon pairs can be created spontaneously from the 
vacuum \cite{Mis93,MaoPre}. 
A schematic picture for this phenomena is depicted in Fig.~1.
The situation is quite similar to the 
electron-positron pair creations in QED with strong electromagnetic
 fields \cite{Gre85}. The produced 
anti-nucleons will then annihilate with the nucleons at the 
ambient medium through 
the $N\bar{N} \rightarrow \gamma\gamma$ reaction. This yields a large
amount of energy and photons. 
This process may happen in addition to the $\nu\bar{\nu}$ annihilation
process.
The sequential process, $\gamma\gamma 
\leftrightarrow e^{+}e^{-}$, inevitably leads to the creation of a fireball.
The dynamical expansion of the fireball will radiate the observed 
$\gamma$-rays through the nonthermal processes in shocks \cite{Pir99}. 
In the following, we shall estimate whether enough energy
is available within this scenario to satisfy the requirement of a source
of energetic GRBs.

We start from the Lagrangian density for nucleons interacting through
the exchange of mesons 
   \begin{eqnarray}
{\cal L}&=&\bar{\psi}[i\gamma_{\mu}\partial^{\mu}-M_{N}]\psi
   + \frac{1}{2}
\partial_{\mu}\sigma\partial^{\mu}\sigma - \frac{1}{2}m_{\sigma}^{2}
 \sigma^{2}
 -\frac{1}{4}\omega_{\mu\nu}\omega^{\mu\nu} \nonumber \\
&& + \frac{1}{2}m_{\omega}^{2}\omega_{\mu}\omega^{\mu}
 - \frac{1}{4} {\bf R}_{\mu\nu}{\bf R}^{\mu\nu}
 +\frac{1}{2}m_{\rho}^{2}{\bf R}_{\mu} \cdot {\bf R}^{\mu}
 \nonumber \\
&& + {\rm g}_{\sigma}\bar{\psi}\psi\sigma
      - {\rm g}_{\omega}\bar{\psi}\gamma_{\mu}\psi\omega^{\mu}
 - \frac{1}{2}{\rm g}_{\rho}\bar{\psi}\gamma_{\mu}\mbox{\boldmath $\tau$}
\cdot \psi {\bf R}^{\mu} ,
    \end{eqnarray}
where the usual notation is used as given in the literature \cite{Ser86}.
Based on this Lagrangian, we have developed a relativistic Hartree approach
including vacuum contributions which describe the properties of nucleons
and anti-nucleons in nuclear matter and finite nuclei
 quite successfully \cite{MaoPre}. The parameters of the model
are fitted to the ground state properties of spherical nuclei. The RHA0 set
of parameters gives 
${\rm g}_{\sigma}^{2}({\rm M}_{N}/m_{\sigma})^{2}=229.67$,
${\rm g}_{\omega}^{2}({\rm M}_{N}/m_{\omega})^{2}=146.31$,
${\rm g}_{\rho}^{2}({\rm M}_{N}/m_{\rho})^{2}=151.90$. It leads to the 
nuclear matter saturation density $\rho_{0}=0.1513$ $fm^{-3}$ 
(0.1484 -- 0.1854 $fm^{-3}$) with a 
binding energy $E_{bind}=-17.39$ MeV ($-16\pm 1$ MeV) 
and a bulk symmetry energy 
$a_{sym}=40.4$ MeV (33.2 MeV). The corresponding 
empirical values are given in parentheses.
The model can be further applied to the neutron-proton-electron
($n$-$p$-$e$) system under the beta equilibrium and the charge
neutrality conditions which is in particular important for the neutron
star. The positive energy of the nucleons in the Fermi sea $E_{+}$ and the
negative energy of the nucleons in the Dirac sea $E_{-}$ can be written as
 \begin{eqnarray}
 E_{+} &=& \left\{ \left[ k^{2} + \left( M_{N} - {\rm g}_{\sigma}\sigma
 \right)^{2} \right] ^{1/2} + {\rm g}_{\omega}\omega_{0}
 + \frac{1}{2}{\rm g}_{\rho}\tau_{0}R_{0,0} \right\}, \\
 E_{-} &=& - \left\{ \left[ k^{2} + \left( M_{N} - {\rm g}_{\sigma}\sigma
 \right)^{2} \right] ^{1/2} - {\rm g}_{\omega}\omega_{0}
 + \frac{1}{2}{\rm g}_{\rho}\tau_{0}R_{0,0} \right\}.    
 \end{eqnarray}
 Here $\sigma$, $\omega_{0}$ and $R_{0,0}$ are the mean values of the scalar
 field, the time-like component of the vector field, and the time-like
 isospin 3-component of the vector-isovector field in neutron star matter,
 respectively. They are obtained by solving the non-linear equations
 of the meson fields including vacuum contributions under the constraints 
 of charge neutrality and general equilibrium. The energy of anti-nucleons
 $\bar{E}_{+}$ is just the negative of $E_{-}$, i.e., $\bar{E}_{+}=
 -E_{-}$ \cite{MaoPre}. By setting $k=0$ in Eqs. (2) and (3), 
 one gets the energies
 of nucleons and anti-nucleons at zero momentum.
 The critical density
 $\rho_{C}$ for nucleon--anti-nucleon pair creation is reached when
 $E_{-}=M_{N}$. 
 The results are given in Fig.~2 where the single-particle energies of the 
 positive-energy nucleon and the negative-energy nucleon are plotted as a 
 function of density.
 Due to the effects of the $\rho$-meson field, $\rho_{C}=6.1$
 $\rho_{0}$ for $p\bar{p}$ pair creation and $7.5$ $\rho_{0}$ for $n\bar{n}$
 pair creation. At the same time, we have calculated the equation
 of state (EOS) of neutron star matter. 
 The structures  and properties of neutron stars  can be 
 obtained by applying the equation of state to solve the Oppenheimer-Volkoff
 equation \cite{Sha83}. The maximum mass of stars turns out to be
 $M_{max}=2.44$ $M_{\odot}$,
 and the corresponding radius $R=12.75$ km and the central density 
 $\rho_{cen}=5.0$ $\rho_{0}$. The $\rho_{cen}$ is smaller than the critical
 density $\rho_{C}$. That means that the spontaneous $N\bar{N}$ pair creation
 does not happen for a  single neutron star within the model employed.

We consider the following case of neutron star collision: Two identical
neutron stars with $\rho_{cen}=4.5$ $\rho_{0}$ (with the current 
EOS, it is related to $M=2.43$ $M_{\odot}$ and $R=13.0$ km)
 collide with each other with a free-fall velocity. 
 The impact parameter $b$ stays between $0$ and $R$,
 which determines the factor of density enhancement. We assume that a 
 compact object of average density $7.2$ $\rho_{0}$ is created in the reaction
 zone. The radius of the compact object is assumed to be
  $r=1$ km (case A) or $r=3$ km
(case B) depending on the values of $b$. Since for a single neutron star with
$\rho_{cen}=4.5$ $\rho_{0}$  the density at $r=1$ km is $4.46$ $\rho_{0}$
and at $r=3$ km is $4.18$ $\rho_{0}$, in case A the density is enhanced
during neutron star collision by a factor around $1.6$ while in case B 
around $1.7$. In both cases the $p\bar{p}$ pair creation will happen while
the contributions of the $n\bar{n}$ pair creation is negligible (it  
contributes at higher density but does not affect our discussions). We 
define a {\em Dirac momentum} $k_{D}$ which describes the negative-energy
nucleons
occupying the eigenstates of the Dirac sea from the {\em uppermost} level
(the lowest-energy antiparticle level) to the negative continuum (see, Fig.~1), 
i.e., $E_{-}=-M_{N}$ in Eq. (3).
At the critical density for $p\bar{p}$ pair creation 
 $\rho_{C}^{p\bar{p}}=6.1$ $\rho_{0}$, the {\em Dirac} momentum 
  $k_{D}^{C}=11.28$ $fm^{-1}$;
and at $\rho=7.2$ $\rho_{0}$, $k_{D}=12.45$ $fm^{-1}$. 
We further define a momentum $p_{max}$ at $E_{-}=M_{N}$, which turns out to be
  \begin{equation}
 p_{max}=\sqrt{ \left( {\rm g}_{\omega}\omega_{0} -\frac{1}{2}{\rm g}_{\rho}
  \tau_{0}R_{0,0} + {\rm g}_{\sigma}\sigma -2M_{N} \right)
  \left( {\rm g}_{\omega}\omega_{0} -\frac{1}{2}{\rm g}_{\rho}\tau_{0}
  R_{0,0} - {\rm g}_{\sigma}\sigma \right) }.
  \end{equation}
Based on the semi-classical phase-space assumption   
we then estimate the number of the $p\bar{p}$ pairs whose energies are
larger than the nucleon free mass 
at $\rho =7.2$ $\rho_{0}$  as
 \begin{equation}
 N_{pair}=\frac{4}{3}\pi r^{3} \times \frac{p_{max}^{3}}{3\pi^{2}} 
  =2.147r^{3} \times 10^{54} .
 \end{equation}
 Before expansion, this compact object remains high density.
Let us check whether most of the $p\bar{p}$ pairs can be created spontaneously.
 The rates for the $N\bar{N}$ pair production per unit surface area and unit 
 time, $dN_{pair}/dSdt$, has been calculated in Ref. \cite{Mis93} for
 compressed matter. In the case of $\rho=7$ $\rho_{0}$ and tunnel distance
 $d=1$ $fm$, the rate turns out to be $2.68\times 10^{-2}$ $fm^{-3}$.
 For case B with $r=3$ $km$, the time needed to emit the available $p\bar{p}$
 pairs is $t=1.9\times 10 ^{19}$ $fm=6.3\times 10^{-5}$ $s$, which is smaller
 than the typical dynamical scale of NS-NS collision $\tau \sim 10^{-3}$ $s$.
Thus, we have enough time to produce proton--anti-proton pairs spontaneously.
The produced protons stay in the atmosphere due to gravitational force. 
However, at that time holes (anti-protons) are still in
bound states due to potentials they feel (a small fraction may be transported
into the negative continuum). The above process may happen before
the compact object expands (we are discussing a microscopic procedure in 
a macroscopic phenomenon). 
Then the compact object expands and the potentials in the 
Dirac sea fall down. Those anti-particles (holes) in bound states are pushed
into the lower continuum and thus escape. They annihilate with the protons
in the atmosphere or in the surrounding objects and release a large sum
of energy.                                             
If one assumes that 80\% of the produced anti-protons annihilate with
protons in the surrounding medium and the released energy is $2$ GeV at each
event (at the moment it's not very clear how many anti-protons in the Dirac-sea
can escape through the lower continuum. This is a problem,
which should be investigated more closely.), 
the total exploded energy $E_{tot}$ turns out to be $5.5 \times
10^{51}$ erg and $1.5 \times 10^{53}$ erg for cases A and B, respectively.
As mentioned before, the efficiency to transfer the initial energy  to the
observed radiation is only 1\% \cite{Kum99}. It seems to be necessary to adopt
the picture of beaming explosion for the most energetic $\gamma$-ray bursts.

Some discussion is now appropriate. Neutron star collisions have repeatedly
been suggested in the literature as possible sources of $\gamma$-ray bursts
\cite{Kat96,Dok96}, powered either by $\nu\bar{\nu}$ annihilation or by highly
relativistic shocks. In Ref. \cite{Ruf98} Ruffert and Janka claimed that a
$\gamma$-ray burst powered by neutrino emission from colliding neutron stars
is ruled out. Here we propose a new scenario caused by the strong 
interactions in dense matter. A large number of anti-particles may be created
from the vacuum when the density is higher than the critical density
for spontaneous particle--anti-particle pair creation. Such high density 
can be reached during the NS-NS collisions, BH-NS mergers, or even NS-NS
mergers when the merged binary neutron stars have large maximum densities.
Some of the produced anti-particles can be ejected from the reaction zone due to
violent dynamics. They may be the novel source of low-energy cosmic-ray
anti-particles which is currently an exciting topic in modern astrophysics
\cite{Mat98}.
Most of them will
annihilate with the corresponding particles
at the ambient medium, and thus release a large amount of energy.
As a first step  we have discussed the $p\bar{p}$ pair creation
in two neutron star collision scenarios because
its critical density is lower than that
of other baryons. Our calculations show that the exploded energy satisfies
the requirement for the initial energy of the energetic GRBs observed
up to now. The variation of the released energies of different GRBs can be 
attributed to the different impact parameters of colliding neutron stars.
Those anti-protons, although produced spontaneously, annihilate during
the dynamical procedure  with  random probability in collisions with protons.
Furthermore, the anti-protons annihilating later might be accelerated by the
photons produced by the nearby $p\bar{p}$ pair annihilations taking place
earlier. This leads to the high-energy anti-protons and, 
consequently, the high-energy
photons. Some of them may escape from the fireball before being distorted by the
medium. Those escaping high-energy photons may
 constitute the observed high-energy tail of $\gamma$-ray bursts.
 This has to be pursued further theoretically. 

In summary, we have proposed a new scenario of vacuum discharge  due to
strong interactions in dense matter as a possible source of 
$\gamma$-ray bursts.
Based on the meson field theoretical model we have estimated the exploded energy
$E_{tot} \sim 10^{51} - 10^{53}$ erg within two neutron star collisions, 
which fits well 
into the range of the initial energy necessary for
 most energetic $\gamma$-ray bursts. For a more quantitative
study, one needs to introduce hyperon degrees of freedom \cite{Gle85} and
even quark degree of freedom \cite{Sch81,Sch99} 
if one assumes that the center of
neutron star is in quark phase. 
Here we have mainly discussed NS-NS collisions. In fact,   
the proposed scenario may happen more  frequently for 
 BH-NS mergers since the production rate for
BH-NS binaries is $\sim$ $10^{-4}$ per yr per galaxy \cite{Bet98}
which is much larger than the rate of direct NS-NS collisions
(for an estimation of collision rate in dense cluster of neutron stars, see
Ref. \cite{Kat96}).
In this case one might obtain a even higher explosion energy reaching the
value of $10^{54}$ erg.
A relativistic  dynamical model
like relativistic  fluid dynamics incorporating meson fields
is highly desirable to simulate
NS-NS collisions, NS-NS/BH-NS mergers. 
At the end, we would like to mention that a similar process may happen in
nucleus-nucleus collisions as discussed in the introduction of 
Ref. \cite{MaoPre} where a dynamical production of anti-matter clusters
due to the variation of the time-dependent meson fields has been 
suggested. We propose to study the photon and anti-proton spectra in
ultra-relativistic heavy-ion collisions which may provide us with
information of structure of discharged vacuum.
Works on these aspects are presently underway.

\end{sloppypar}

 \vspace{0.5cm}
 \begin{sloppypar}
 \noindent {\bf Acknowledgements:} 
 The authors thank N.K.~Glendenning for fruitful comments on the preliminary
 version of manuscript.
 G.~Mao acknowledges  the STA  foundation                       
 for financial support      
 and  the members of the Research Group for Hadron Science 
 at Japan Atomic Energy Research Institute
 for their hospitality.

 \end{sloppypar}

\newpage
 \begin{figure}[htbp]
  \vspace{0cm}
 \hskip  -0.5cm \psfig{file=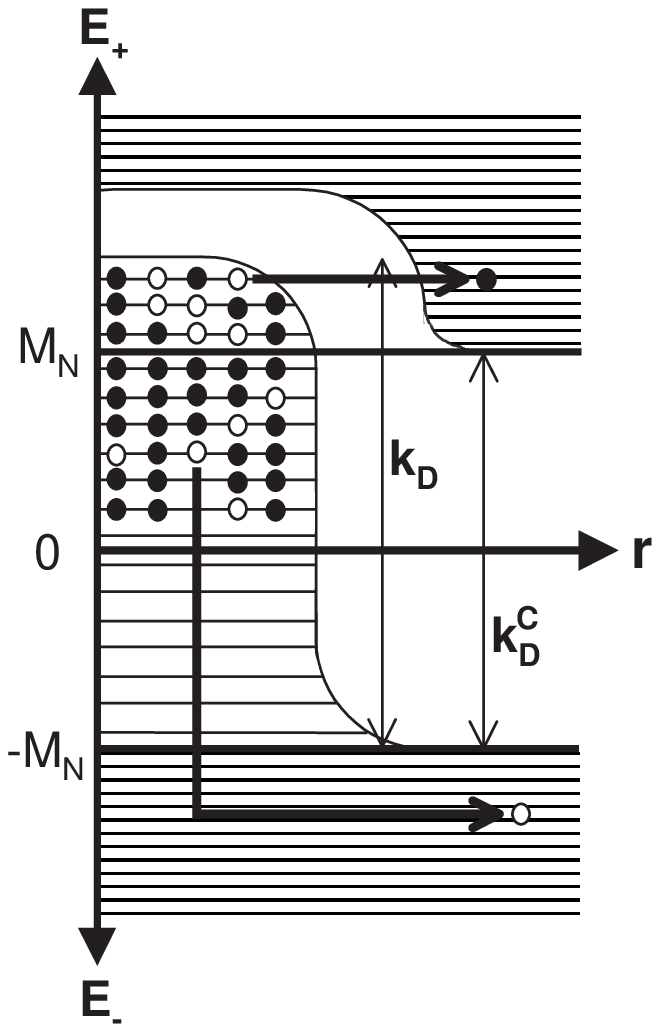,width=15.cm,height=17cm,angle=0}
 \caption{Schematic view of $N\bar{N}$ pair creation 
 from the Dirac sea due to 
 strong fields in dense matter.}
\end{figure}
 \newpage
 \begin{figure}[htbp]
  \vspace{0cm}
 \hskip  -0.5cm \psfig{file=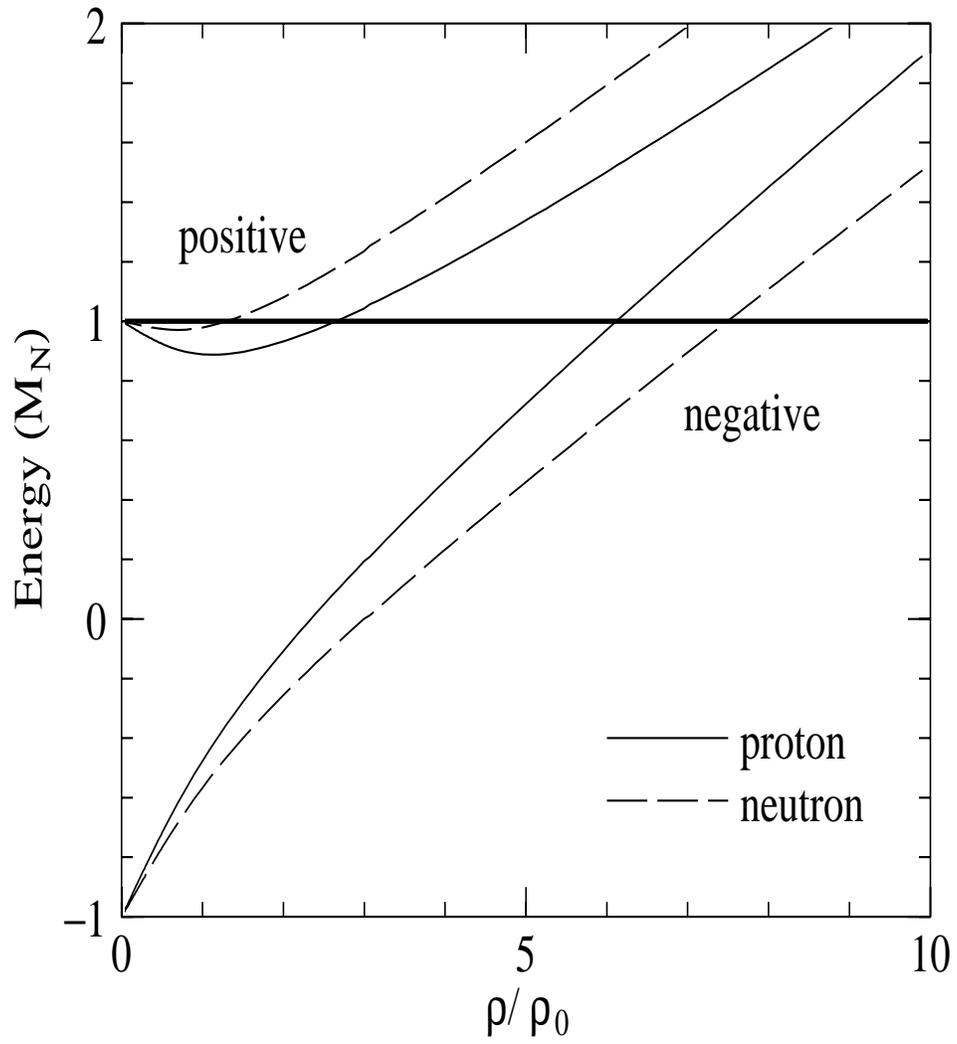,width=15.cm,height=17cm,angle=0}
 \caption{The single-particle energies of the positive-energy nucleon and 
 the negative-energy nucleon
 in neutron star matter.}
\end{figure}

\end{document}